\documentclass[preprint,aps,nofootinbib]{revtex4-1}

\usepackage{amsmath}  
\usepackage{amsbsy,amssymb,amsfonts}
\usepackage{graphicx}
\usepackage{color}
\usepackage{epsf}
\usepackage[utf8x]{inputenc}
\usepackage{hyperref}
\usepackage{multirow}
\usepackage{ulem}
\def\bravert{\egroup\,\vrule\,\bgroup}

\newcommand{\beq}{\begin{equation}}
\newcommand{\eeq}{\end{equation}}
\newcommand{\beqa}{\begin{eqnarray}}
\newcommand{\eeqa}{\end{eqnarray}}
\newcommand{\bea}{\begin{array}}
\newcommand{\eea}{\end{array}}

\newcommand{\bef}{\begin{figure}}
\newcommand{\ef}{\end{figure}}
\newcommand{\bc}{\begin{center}}
\newcommand{\ec}{\end{center}}
\newcommand{\bt}{\begin{table}}
\newcommand{\et}{\end{table}}
\newcommand{\btb}{\begin{tabular}}
\newcommand{\etb}{\end{tabular}}

\def\rvac{\left| \rule{0.3cm}{.0cm} \right>}

\newcommand{\ident}{1\!\!1}
\newcommand{\ols}[1]{\mskip.5\thinmuskip\overline{\mskip-.5\thinmuskip {#1} \mskip-.5\thinmuskip}\mskip.5\thinmuskip} 

\newcommand{\olsi}[1]{\,\overline{\!{#1}}} 
\makeatletter
\newcommand\closure[1]{
	\tctestifnum{\count@stringtoks{#1}>1} 
	{\ols{#1}} 
	{\olsi{#1}} 
}
\long\def\count@stringtoks#1{\tc@earg\count@toks{\string#1}}
\long\def\count@toks#1{\the\numexpr-1\count@@toks#1.\tc@endcnt}
\long\def\count@@toks#1#2\tc@endcnt{+1\tc@ifempty{#2}{\relax}{\count@@toks#2\tc@endcnt}}
\def\tc@ifempty#1{\tc@testxifx{\expandafter\relax\detokenize{#1}\relax}}
\long\def\tc@earg#1#2{\expandafter#1\expandafter{#2}}
\long\def\tctestifnum#1{\tctestifcon{\ifnum#1\relax}}
\long\def\tctestifcon#1{#1\expandafter\tc@exfirst\else\expandafter\tc@exsecond\fi}
\long\def\tc@testxifx{\tc@earg\tctestifx}
\long\def\tctestifx#1{\tctestifcon{\ifx#1}}
\long\def\tc@exfirst#1#2{#1}
\long\def\tc@exsecond#1#2{#2}
\makeatother

\makeatletter

\newcommand{\pushright}[1]{\ifmeasuring@#1\else\omit\hfill$\displaystyle#1$\fi\ignorespaces}
\newcommand{\pushleft}[1]{\ifmeasuring@#1\else\omit$\displaystyle#1$\hfill\fi\ignorespaces}
\makeatother

\begin{document}

\title{Semi-Hadronic Charge-Parity Violation Interaction Constants in CsAg, FrLi and FrAg molecules}

\vspace*{1cm}

\author{Timo Fleig}
\email{timo.fleig@irsamc.ups-tlse.fr}
\affiliation{Laboratoire de Chimie et Physique Quantiques,
             FeRMI, Universit{\'e} Paul Sabatier Toulouse III,
             118 Route de Narbonne, 
             F-31062 Toulouse, France }

\author{Aur\'elien Marc}
\email{marc.aurelien.am@gmail.com}
\affiliation{Laboratoire de Chimie et Physique Quantiques,
	FeRMI, Universit{\'e} Paul Sabatier Toulouse III,
	118 Route de Narbonne,
	F-31062 Toulouse, France}
	
\author{Timo Fleig}
\email{timo.fleig@irsamc.ups-tlse.fr}
\affiliation{Laboratoire de Chimie et Physique Quantiques,
	FeRMI, Universit{\'e} Paul Sabatier Toulouse III,
	118 Route de Narbonne, 
	F-31062 Toulouse, France }
\vspace*{1cm}
\date{\today}

\bigskip

\begin{abstract}
We present a systematic study of the nucleon-electron tensor-pseudotensor (Ne-TPT) interaction in candidate molecules for 
next-generation experimental searches for new sources of charge-parity violation. 
The considered molecules are all amenable to assembly from laser-cooled atoms, with the francium-silver (FrAg) molecule 
previously shown to be the most sensitive to the Schiff moment interaction in this set.
Interelectron correlation effects are treated through relativistic general-excitation-rank configuration-interaction theory in
the framework of the Dirac-Coulomb Hamiltonian.
We find in FrAg the Ne-TPT interaction constant to be $W_T({\text{Fr}}) = 2.58 \pm 0.21 [\left<\Sigma\right>_A \mathrm{kHz}]$,
considering the Francium atom as target of the measurement.
Taking into account nuclear structure in a multi-source interpretation of a measured electric dipole moment, FrAg is found to 
be an excellent probe of physics beyond the Standard Model as this system will in addition to its sizeable Ne-TPT
interaction constant greatly constrain fundamental 
parameters such as the quantum-chromo-dynamic $\bar{\theta}$ or the semileptonic four-fermion interaction $C_{lequ}$ from 
which nuclear and atomic ${\cal{CP}}$-violating properties arise.
\end{abstract}

\maketitle
\section{Introduction}
\label{SEC:INTRO}
A lacking explanation for the baryon asymmetry of the Universe (BAU) is one of the greatest shortcomings of the standard model (SM) of particle physics. 
Even though some of this asymmetry can be explained within the SM through the Cabibbo-Kobayashi-Maskawa (CKM) formalism 
\cite{Cabibbo_1963,Kobayashi_Maskawa_1973}, it is not sufficient to explain the observed BAU 
\cite{EWBG_MRM2012,EWBG_Sather_1994}.
Then, electric dipole moments (EDM) serve as powerful low-energy probes \cite{EDM_new_phi_2022} in the quest for new sources of charge-parity ($\mathcal{CP}$) violation. Atoms and molecules, with their complex structures, offer significant advantages in this pursuit as they can amplify new sources of $\mathcal{CP}$ violation by several orders of magnitude \cite{sandars_atomicEDM1968,new_phy_atom_and_mol_2018}. 
However, the complexity of those systems introduces numerous potential underlying $\mathcal{CP}$-violating mechanisms at the 
nuclear and atomic scales, necessitating multiple measurements across different systems to distinguish the possible sources \cite{Chupp_Ramsey_Global2015}. 
In carefully selected atoms and molecules, leptonic $\mathcal{CP}$ violation can be significantly suppressed, rendering these systems primarily sensitive to hadronic and specific semihadronic sources \cite{Barr_1992}. 

In a recently published paper \cite{FrAg_Schiff} some of us studied the Schiff moment interaction in different diatomic molecules constituted of laser-coolable atoms. Nevertheless, as the Schiff moment interaction is not the only unsuppressed possible source of $\mathcal{CP}$-violation, other calculations on the different contributions are required. We thus present here a systematic study of nucleon-electron tensor-pseudotensor (Ne-TPT) interaction in some of the previously studied molecules, Francium-Silver (FrAg), Francium-Lithium (FrLi) and Cesium-Silver (CsAg). 
Francium (Fr) and Cesium (Cs) serve as highly polarizable target atoms while Lithium (Li) and Silver (Ag) are polarizing
partner atoms.
We emphasize those two specific target atoms since Fr has the biggest Schiff moment interaction constant ($W_S$). Using
Cs as a target atom is less favorable in this respect, but since long-lived isotopes ($^{133}\mathrm{Cs}$) exist
possible experiments with Cs-containing molecules may be easier to conduct than with molecules containing radioactive 
francium nuclei. Following similar arguments, proposals have been made concerning alternatives to the most favorable --- 
in regard of its
$\mathcal{CP}$-odd interaction constants --- but radioactive Radium-Silver (RaAg) molecule \cite{Fleig_DeMille_2021} 
in electron EDM searches. The molecules Ytterbium-Copper (YbCu),
Ytterbium-Silver (YbAg) \cite{Vutha,Tomza_YbAg} or Barium-Silver (BaAg) \cite{Dietrich} can all be built from long-lived
nuclear isotopes and may be used as stable alternatives or precursors to RaAg experiments.
Concerning the partner atoms, Li mainly serves for comparative purposes with Ag and to establish better qualitative comprehension of the physics occurring in such complex systems. 
The main purpose of this paper is thus to present our results on the Ne-TPT interaction in the three presented Alkali-{metal}-Alkali-{metal} molecules.
We will make the case that the FrAg molecule is both highly sensitive to the Ne-TPT interaction as well as a favorable system
for more strongly constraining ${\cal{CP}}$-violating parameters in global analyses of the EDM landscape 
\cite{Chupp_Ramsey_Global2015,Degenkolb_EDMlandscape2024}.

The present work is structured as follows. In \autoref{SEC:THEORY} we lay out the molecular theory for obtaining the wave functions and how they are used to compute the Ne-TPT molecular interaction constant.
The main aspects of our work are presented in \autoref{SEC:APPL} where we present and discuss the results for $W_T$ in the three molecules of interest. 
Finally in \autoref{SEC:CONCL} we discuss the obtained values and their possible impact in the search for $\mathcal{CP}$-violation beyond SM.
\section{Theory}
\label{SEC:THEORY}
As extensively explained in Refs. \cite{Fleig_Jung_Xe_2021,NeTPT_mol_fleig2024} the energy shift due to the Ne-TPT interaction is given by:
\begin{equation}
	\Delta\varepsilon_{^\mathrm{TPT}} = \left< \psi \right| \hat{H}^{\mathrm{eff}}_{\mathrm{TPT}} \left| \psi \right> = W_T C_T^A
\end{equation}
where $C^A_T$ is a $\mathcal{CP}$-odd parameter that is isotope specific and reflects the semi-hadronic character
of the Ne-TPT interaction, and $A$ is the nucleon number. The molecular Ne-TPT interaction constant is expressed as:
\begin{equation}
	W_T(X) = \sqrt{2}G_F \left< \Sigma \right>_A 
	\left< \psi \right| 
	\imath \sum_{j=1}^n (\gamma_3)_j \rho_X({\bf r}_j)
	\left| \psi \right>
   \label{EQ:NETPT}
\end{equation}
where $X$ denotes the nucleus of interest, $\rho_X({\bf r})$ its density at position ${\bf r}$, and $\gamma_3$ is the standard 
Dirac matrix aligned with the $z$-axis since the molecule is aligned along this specific axis. Finally, $\left< \Sigma \right>_A$ 
is the nuclear-spin expectation value for a given isotope. 
The Dirac matrix $(\gamma_3)_j$ for electron $j$ in the above operator goes to unity in the non-relativistic limit of
the theory. Therefore, Ne-TPT interactions are intrinsically relativistic in origin which is why the use of heavy target atoms
is indicated where electrons achieve mean velocities a significant fraction of the speed of light.
The expectation value in Eq. (\ref{EQ:NETPT}) is evaluated over the molecular wave function $\psi$ obtained from a 
configuration-interaction (CI) expansion of the zeroth order eigenvalue problem:
\begin{equation}
	\hat{H}^{DC} \psi = \varepsilon \psi
\end{equation}
with $\hat{H}^{DC}$ the Dirac-Coulomb Hamiltonian. In a diatomic molecule with $n$ total electrons, it reads as:
\begin{equation}
	\hat{H}^{DC}= \sum_{j=1}^{n} \left[ c {\boldsymbol \alpha}_j \cdot {\bf p}_j + \beta_j c^2 - \sum_K^2\frac{Z_K}{r_{jK}} \ident_4 \right] + \frac{1}{2} \sum_{j,k}^{n} \frac{1}{r_{jk}}\ident_4 + V_{KL}
\end{equation}
where ${\boldsymbol \alpha}$ and $\beta$ are electronic Dirac matrices, $K$ runs over the two nuclei and $V_{KL}$ is the electrostatic potential energy for two fixed nuclei in the Born-Oppenheimer approximation.
The CI wave function is expanded as: 
\begin{equation}
	\left| \psi \right> = \sum_{I=1}^{dim\mathcal{F}^t(M,N)} c_{(M_J),I} \hat{\mathcal{T}}{}^\dagger \hat{\closure{\mathcal{T}}}{}^\dagger \rvac
\end{equation}
where $\rvac$ is the true vacuum state, $\mathcal{F}^t(M,N)$ is the symmetry restricted sector of Fock space ($M_J$ subspace) with $n$ electrons in $M$ four-spinors, $\hat{\mathcal{T}} = \hat{a}_i^\dagger \hat{a}_j^\dagger ...$ is a string of spinor creation operators, $ \hat{\closure{\mathcal{T}}}{}^\dagger = \hat{\closure{a}}_k^\dagger \hat{\closure{a}}_l^\dagger ...$  is a string of creation operators of Kramers-partner spinors, and $c_{(M_J),I}$ are the determinant expansion coefficients obtained with the procedure given in Refs. \cite{GAS_2001,GAS_2003}.
\section{Applications}
\label{SEC:APPL}
\subsection{Technical details}
In this work, the target atoms (Fr, Cs) and their partners (Ag, Li) are described by the same basis sets as presented in Ref. 
\cite{FrAg_Schiff}. To summarize for the target atoms, 
we use Dyall's quadruple-zeta (QZ) basis sets
\cite{Dyall_2009}. When those sets are densified, they are denoted as QZ+. For comparative purpose, the double- and triple-zeta 
(DZ and TZ resp.) Dyall's basis sets are used for the Fr atom.

All numerical calculations were carried out using a locally modified version of the DIRAC program package \cite{DIRAC_code}. 
In general, a Dirac-Coulomb-Hartree-Fock (DCHF) calculation yields the set of molecular spinors with which the configuration-interaction (CI) expansion is performed.
For all correlated calculations we use the KRCI module \cite{luciarel_parallel} and
the Generalized Active Space (GAS) \cite{GAS_2001} formalism allowing for general types of excitations in the CI models.
The 
notation used for the different models stands as
$\mathrm{S}i\_\mathrm{SD}j\_\mathrm{SDT}k\_\mathrm{SDTQ}l\_space$.
One should read:
\begin{itemize}
	\item $i$ electrons are in shells from which single excitations are performed
	\item $j$ electrons are in accumulated shells with single and double excitations
	\item $k$ electrons are in accumulated shells with up to triple excitations
	\item $l$ electrons are in accumulated shells with up to quadruple excitations 
\end{itemize}
$space$ denotes the complementary space. In most cases, it is a number that indicates the cutoff in atomic units in the complementary spinor space, deleting all functions with higher energy than the one indicated in the cutoff and keeping all the ones with lower energy. In one specific case in this work the complementary space is ``cc" standing for ``customized cutoff" that will be explained in its dedicated section.

\subsection{Calculations}

In this section we present results for the three molecules as data tables where the model, the total energy, and the Ne-TPT interaction in two different units are given. 
The total energies we obtained are coherent with their respective models. 
In particular, it can be seen that correlation energies per electron are in the expected range of 0.01a.u. to 0.03a.u.

\subsubsection{Basis set analysis}

Here we justify the use of a non-densified QZ basis set instead of the densified QZ+ one like in our precedent work on the Schiff moment interaction \cite{FrAg_Schiff}.
A densified basis set concerns the target atom and is obtained through a densification procedure \cite{Hubert_Fleig_2022} of a given basis set improving the description of the atomic shells but leading to greater computational demand. 
With the help of Table \ref{Table: FrAg basis ana}, one can see that, at the DCHF level, the convergence of $W_T(\mathrm{Fr})$ is already achieved using the QZ basis set, within a 0.1\% error margin. 
Indeed, the error compared to the densified basis set is roughly 0.03\%. 
When considering single-double excitations from the valence shell, which is the most contributing shell as we show later in this work, the difference between QZ and QZ+ basis sets remains at about 0.03\%.
Meanwhile, going from DZ to TZ basis sets makes a difference of the percent order for both DCHF and SD2 models.
From TZ to QZ basis the difference at DCHF level is of 0.06\% and 0.1\% when considering the SD2\_8a.u. model. 
\begin{table}[h] 
	\centering
	\caption{
		FrAg, $^1\Sigma_0$, $R_e =$ 6.190 a.u. 
}
	\label{Table: FrAg basis ana}
	\begin{tabular}{l|c|c|r}
		Basis/cutoff            &  $\varepsilon_{\text{CI}}$ [a.u.] & $W_T(\mathrm{Fr})$ [$10^{-13} \left<\Sigma\right>_A$ a.u.] & $W_T(\mathrm{Fr})$ [$\left<\Sigma\right>_A$ kHz]  \\
		\hline
		cvDZ/DCHF              & -29622.8235344 & 4.11
		& 2.70\\
		cvDZ/SD2\_8a.u.        & -29622.8485339 & 3.97
		& 2.61\\
		\hline
		cvTZ/DCHF              & -29622.8356809 & 4.15
		& 2.73\\
		cvTZ/SD2\_8a.u.        & -29622.8607613 & 4.01
		& 2.64\\
		\hline
		cvQZ/DCHF	           & -29622.8363746 & 4.15
		& 2.73\\
		cvQZ/SD2\_8a.u.        & -29622.8615002 & 4.01
		& 2.64\\
		\hline
		cvQZ+/DCHF	           & -29622.8354238 & 4.15
		& 2.73\\
		cvQZ+/SD2\_8a.u.       & -29622.8605500 & 4.01
		& 2.64\\
	\end{tabular}
\end{table}

Therefore, we use the QZ basis set for the FrAg molecule for computational efficiency.
In the case of the FrLi and CsAg molecules we use the QZ+ basis sets due to already in hand results and spinors from our previous study. 
As a general rule, we observe that increasing the extent of the atomic basis sets increases the Ne-TPT interaction. This is
reasonable since the bulk of the atomic closed shells contributes to the interaction constant and a more extensive basis set will
improve the description of all these shells.

\subsubsection{Francium target atom}

In Tables \ref{Table : FrLi W_T} and \ref{Table : FrAg W_T} our calculations for the two diatomic molecules FrLi and FrAg are shown. 
They display the obtained energies and the Ne-TPT interaction constant $W_T(\mathrm{Fr})$ from various CI models.

\begin{table}[h] 
	\centering
	\caption[]{FrLi, $^1\Sigma_0$, $R_e =$ 6.878 a.u.
	}
	\label{Table : FrLi W_T}
	\begin{tabular}{l|c|c|r}
		Basis/cutoff            &  $\varepsilon_{\text{CI}}$ [a.u.] & $W_T(\mathrm{Fr})$ [$10^{-13} \left<\Sigma\right>_A$ a.u.] & $W_T(\mathrm{Fr})$ [$\left<\Sigma\right>_A$ kHz]  \\
		\hline
		cvQZ+/DCHF            & -24315.6237749 & 3.40
		& 2.23\\ 
		cvQZ+/SD2\_10a.u.     & -24315.6516000 & 3.23
		& 2.13\\ 
		cvQZ+/SD10\_10a.u.    & -24315.8126807 & 3.28
		& 2.16\\
		cvQZ+/SD22\_10a.u.    & -24316.2122877 & 3.29
		& 2.16\\
	\end{tabular}
\end{table}

\begin{table}[h] 
	\centering
	\caption[]{FrAg, $^1\Sigma_0$, $R_e =$ 6.190 a.u. 
	}
	\label{Table : FrAg W_T}
	\begin{tabular}{l|c|c|r}
		Basis/cutoff            &  $\varepsilon_{\text{CI}}$ [a.u.] & $W_T(\mathrm{Fr})$ [$10^{-13} \left<\Sigma\right>_A$ a.u.] & $W_T(\mathrm{Fr})$ [$\left<\Sigma\right>_A$ kHz]  \\
		\hline
		cvQZ/DCHF	           & -29622.8363746 & 4.15
		& 2.73\\
		cvQZ/SD2\_8a.u.        & -29622.8615002 & 4.01
		& 2.64\\
		cvQZ/SD10\_3a.u.       &                & 4.04
		& 2.66\\
    	cvQZ/SD10\_cc          &                & 4.01
    	& 2.64\\
		cvQZ/SD10\_8a.u.       & -29623.0206732 & 4.02
		& 2.64\\
		cvQZ/SD10\_11.5a.u.    &                & 4.01
        & 2.64\\
		cvQZ/SD8\_SDT10\_3a.u. &                & 4.04
		& 2.66\\
		cvQZ/SD8\_SDTQ10\_3a.u.&                & 3.94
		& 2.60\\
		cvQZ/SD8\_SDT10\_cc    &                & 4.02
		& 2.64\\
		cvQZ/SD12\_8a.u.       & -29623.1930035 & 4.04
		& 2.66\\
		cvQZ/SD20\_8a.u.       & -29623.3380747 & 4.01
		& 2.63\\
		cvQZ/SD36\_8a.u.       & -29623.8389564 & 4.02
		& 2.64\\
	\end{tabular}
\end{table}

Firstly, we compare the two diatomic molecules.
At the DCHF level there is roughly a factor of $1.2$ difference between $W_T^\mathrm{FrLi}(\mathrm{Fr})$ and 
$W_T^\mathrm{FrAg}(\mathrm{Fr})$. This phenomenon has also been observed for the Schiff moment interaction
in \cite{FrAg_Schiff}. 
The explanation is thus very similar and largely relies on the difference in electron affinity between Li and Ag \cite{Li_EA,Ag_EA}.  
Secondly, one can see that in both molecules the most contributing shell is the valence shell formed of the $7s$ electron from Fr and the $ns$ electron of the partner atom ($n = 5$ for Ag and $n = 2$ for Li).
For the FrAg molecule, single and double excitations from the valence shell lower $W_T^\mathrm{FrAg}(\mathrm{Fr, DCHF})$ by 
3.3\% while in the FrLi molecule the lowering is roughly 5\%.

Differently from FrAg, in FrLi more correlated models always increase the $W_T(\mathrm{Fr})$ value. Indeed, compared to SD2, 
the SD10 model increases its value by 1.5\% and adding the $5d$(Fr) and $1s$(Li) shells (SD22\_10a.u model) increases 
$W_T$ by 0.2\%. Overall the SD22\_10a.u. model diminishes the DCHF value by 3.2\%.

Nevertheless, in the FrAg diatomic molecule, the global diminishing from DCHF to SD36 is also 3.2\%. This preserves the factor of
$1.2$, showing that Ag as an atom partner leads to an increased Ne-TPT interaction constant.
In the following we study more closely the FrAg diatomic molecule since it yields a significantly greater interaction constant.

The SD10\_8a.u. model adds contributions from $6s6p$(Fr) to the valence-shell contributions and increases the value of 
$W_T^\mathrm{FrAg}(\mathrm{Fr, SD2\_8a.u.})$ by 0.05\%. Adding the $4d$(Ag) shell to the valence-shell (SD12\_8a.u.) model 
increases the interaction constant value by 0.9\%. Nevertheless, adding both contributions from $6s6p$(Fr) and $4d$(Ag) shells 
to the valence-shell (SD20\_8a.u) model lowers the value by 0.2\%. This result indicates that subtle physics occurs here. 
Indeed, when taken individually the single and double excitations in concerned shells increase the value. 
This evidence, combined with the fact that we expect from theory the $6s6p$(Fr) shells to give greater contributions to 
the $W_T^\mathrm{FrAg}(\mathrm{Fr})$ after the valence shell, is a motivation 
to study higher-excitation-rank models in the valence shell with SD8\_SDT(Q)10 models.

SD8\_SDT(Q)10 models are very demanding in computational resources. Hence, a modification of the cutoff is required. 
In \autoref{Table : FrAg W_T} the SD10\_\textit{space} lines are of interest to justify the cutoffs used. We mainly use a cutoff at 8 atomic units in the virtual space, leading to a virtual space composed of 204 functions.
The 11.5a.u. cutoff corresponds to an additional 95 functions and corrects $W_T$ by roughly 0.04\%. 
This shows that there is no benefit with calculations using this cutoff due to a marginal effect coming with great
additional costs in resources. 
The ``cc" cutoff stands for ``customized cutoff" where a selection of virtual orbitals was performed. We applied a 5 a.u. cutoff 
while excluding a Ag $g$-type shell
since we expect this type of shell to contribute negligibly due to a high principal quantum number. 
This customized cutoff is robust since the results differ from those with the 8 a.u. model by only 0.05\% (which fortuitously 
brings the result closer to the higher 11.5 a.u. cutoff) and is used for the SD8\_SDT10 calculation. 
Nevertheless, this setup is still too demanding for quadruple excitations and a 3 a.u. cutoff is applied to carry out this 
higher level of correlated calculation. The 3 a.u cutoff describes the physics required for $W_T$ to a somewhat lesser degree,
but we still consider such a model as quantitatively acceptable regarding the effect of quadruple excitations, as to be seen 
later.

Now we focus on the effect of triple and quadruple excitations as described by the SD8\_SDT(Q)10 models.
On the one hand, including 
triple excitations while correlating the valence shell increases the value of $W_T$ by 0.2\% using the customized cutoff 
and by 0.1\% using the 3 a.u. cutoff. 
On the other hand, quadruples are seen to have a greater effect since the reduction from SD10\_3a.u. is about 2.4\%. 
This confirms our earlier findings on the importance of higher excitation ranks in a previously studied molecule \cite{NeTPT_mol_fleig2024}.

Finally, the SD36\_8a.u. model correlates electrons from $5d$(Fr) and $4p$(Ag) shells. The resulting increase, compared to SD20, is 0.3\% which finally yields an overall diminution of 3.2\% with respect to the DCHF result. 

For obtaining the final value of $W^\mathrm{FrAg}_T(\mathrm{Fr})$ we use the result from the SD36 
model as base value and add corrections from triple and quadruple excitations as follows:
\begin{equation}
	\begin{aligned}
		W^\mathrm{FrAg}_T(\mathrm{Fr}) &= 
		W^\mathrm{FrAg}_T(\mathrm{Fr, SD36\_8a.u.}) \\
		&-
		W^\mathrm{FrAg}_T(\mathrm{Fr, SD10\_cc}) + 
		W^\mathrm{FrAg}_T(\mathrm{Fr, SD8\_SDT10\_cc})\\
		&- 
		W^\mathrm{FrAg}_T(\mathrm{Fr, SD8\_SDT10\_3a.u.}) + 
		W^\mathrm{FrAg}_T(\mathrm{Fr, SD8\_SDTQ10\_3a.u.}) \\
		W^\mathrm{FrAg}_T(\mathrm{Fr})
		&= 3.93 [10^{-13} \left<\Sigma\right>_A \mathrm{a.u.}] = 2.58 [\left<\Sigma\right>_A \mathrm{kHz}]
	\end{aligned}
\end{equation}
To this final value we attribute an uncertainty of 8\% which is justified as follows.
The employed Dirac-Coulomb Hamiltonian lacks Breit and other higher order terms. For ${\cal{CP}}$-odd interaction
constants this approximation is on the order of $1-2$\% \cite{Skripnikov_HfF+_JCP2017}. The remaining six parts correspond 
to the numerical description of the molecule including basis set, excitation rank, number of correlated electrons and cutoff 
in the virtual space. Including uncertainties our result for the Ne-TPT interaction constant in FrAg is
$W_T^\mathrm{FrAg}(\mathrm{Fr}) = {3.93} \pm {0.32} [10^{-13} \left<\Sigma\right>_A \mathrm{a.u.}] = {2.58} \pm {0.21} [\left<\Sigma\right>_A \mathrm{kHz}]$.

\subsubsection{Cesium target atom}

In \autoref{Table : CsAg W_T} are referenced the results for the CsAg molecule. As in FrAg the valence electrons have the most 
important effect on the $W_T$ lowering, accounting for 85\% of the global diminution \textit{id est} using the most correlated 
model SD20. Adding the $4d$(Ag) shell to the valence model (giving the SD12 model) increases the interaction constant by less than one 
percent, while adding the $5s5p$(Cs) shells diminishes the value by 1.3\%. The overall change between the DCHF level and the 
most correlated model is of the order of 3.2\%.
Thus global trends are the same as for FrAg, especially the correlation effects from the $4d$(Ag) shell and those from
$5s5p$(Cs) which are similar to the correlation effects of the $6s6p$(Fr) electrons in the FrAg molecule.

\begin{table}[h] 
	\centering
	\caption{CsAg, $^1\Sigma_0$, $R_e =$ 6.878 a.u.
	}
	\label{Table : CsAg W_T}
	\begin{tabular}{l|c|c|r}
		Basis/cutoff            &  $\varepsilon_{\text{CI}}$ [a.u.] & $W_T(\mathrm{Cs})$ [$10^{-13} \left<\Sigma\right>_A$ a.u.] & $W_T(\mathrm{Cs})$ [$\left<\Sigma\right>_A$ kHz]  \\
		\hline
		cvQZ+/DCHF           & -13101.4159636 & 0.396
		& 0.254\\
		cvQZ+/SD2\_8a.u.     & -13101.4419586 & 0.376
		& 0.247\\
		cvQZ+/SD12\_8a.u.    & -13101.8080428 & 0.379
		& 0.249\\
		cvQZ+/SD20\_8a.u.    & -13101.9680564 & 0.374
		& 0.246\\
	\end{tabular}
\end{table}

We estimate an uncertainty of 10\% for $W_T^\mathrm{CsAg}(\mathrm{Cs,SD20\_8a.u.})$ by analogy with the FrAg molecule. 
As for the latter, 2 parts come from Hamiltonian approximations. The rest is ascribed to the numerical description of the molecule.
These considerations bring the retained value for the Cesium Ne-TPT interaction constant to
$W_T^\mathrm{CsAg}(\mathrm{Cs,SD20\_8a.u.})=0.374\pm 0.030\cdot 10^{-13} \left<\Sigma\right>_A \mathrm{a.u.} = 0.246\pm 0.025 \left<\Sigma\right>_A\mathrm{kHz}$. 
$W_T^\mathrm{CsAg}(\mathrm{Cs})$ is about one order of magnitude smaller than $W_T^\mathrm{FrAg}(\mathrm{Fr})$ but 
this drawback may be attenuated by the use of a stable target-atom nucleus.

\section{Conclusion and Prospects}
\label{SEC:CONCL}

To extract information about parity and time violating parameters from the molecular EDM, we computed the Ne-TPT interaction constant in three molecules of interest for the next generation of experiments \cite{DeMille_Talk_Seattle}. The results, summarized in \autoref{Table:final}, were obtained with a high level of electron correlation through the CI method.  
\begin{table}[h]
	\centering
	\caption{Summary table of the computed $W_T$ central values\footnote{
			For uncertainties, refer to the dedicated section of the paper.}. $A$ stands for the target nucleus.
	}
	\label{Table:final}
	\begin{tabular}{l|c|c|r}
		Molecule            &  Model & $W_T(A)$ [$10^{-13} \left<\Sigma\right>_A$ a.u.] & $W_T(A)$ [$\left<\Sigma\right>_A$ kHz]  \\
		\hline
		FrAg & final              & 3.93  & 2.58 \\
		FrLi & cvQZ+/SD22\_10a.u. & 3.29  & 2.16 \\
		CsAg & cvQZ+/SD20\_8a.u.  & 0.374 & 0.246 \\
	\end{tabular}
\end{table}
We show that Silver as atom partner is more effective than Lithium and can enhance $W_T$ roughly by 20\% when used in place of the latter. 
Nevertheless, compared to $W_T^\mathrm{TlF}(\mathrm{Tl})$\cite{NeTPT_mol_fleig2024} in the TlF molecule with experiment currently ongoing \cite{CeNTREX_2021}, $W_T^\mathrm{FrAg}(\mathrm{Fr})$ is 37\% weaker disregarding isotope-dependent nuclear spins. 
This reduction is comparable to what we found in the previous Schiff moment interaction constant ($W_{SM}$) study \cite{FrAg_Schiff}. 
The explanations are identical, the partial negative charge on F is greater in TlF than the one on Ag in FrAg. 
The same is true for the $s-p$ mixing which is greater in TlF than in FrAg referring to a DCHF calculation and a Mulliken Analysis of the Gross populations. 
Furthermore, correlating electrons reduces $W^\mathrm{TlF}_{SM}(\mathrm{Tl})$\cite{Hubert_Fleig_2022} and $W^\mathrm{TlF}_T(\mathrm{Tl})$\cite{NeTPT_mol_fleig2024} by 7\% and 12\% resp. while $W^\mathrm{FrAg}_{SM}(\mathrm{Fr})$\cite{FrAg_Schiff} and $W^\mathrm{FrAg}_T(\mathrm{Fr})$ are reduced by 4\% and {5.3\%} resp.
This reduction is also explained by the nature of the chemical bond, different in both molecules. Concerning TlF, the bonding strongly mixes $s$ and $p$ shells from Tl and F resp. Therefore, including excitations reduces the mixing of those shells, leading to a lowering of the considered interaction. 

To study the impact of the Schiff moment or the Ne-TPT interaction on global fits on $\mathcal{CP}$-violating parameters, we 
introduce a ratio {${\tilde{M}}=W_{SM} / W_T$ of} those two constants.
\begin{table}
	\centering
	\caption[]{Ratio for relative suppression, ${}^*$ stands for present work results}
	\label{Table:Ratio_SM/T}
	\begin{tabular}{l|c|c|c|c|c}
		System & $W_{SM}$ [a.u.] & $W_T$ [$10^{-13} \left<\Sigma\right>_A$ a.u.] 
                     & ${\tilde{M}}$ [$10^{13} \left<\Sigma\right>_A^{-1}$ {a.u.}] 
                     & $I_A$ \cite{Fr_spin_expect_val,Tl_spin_expect_val,Cs_spin_expect_val} & $M$ [$10^{13}$ {a.u.}] \\ \hline
   {$^{223}$Fr}Ag  & 30168\cite{FrAg_Schiff} & {$3.93^*$}                      & $7676$  &  $3/2$   &  $5117$    \\
   {$^{133}$Cs}Ag & 3529.6\cite{FrAg_Schiff} & $0.374^*$                      & $9437$  &  $7/2$   &  $2697$    \\
   {$^{205}$Tl}F  & 40539\cite{CeNTREX_2021} & 6.25\cite{NeTPT_mol_fleig2024} & $6486$  &  $1/2$   &  $12972$ 
	\end{tabular}
\end{table}
\autoref{Table:Ratio_SM/T} shows that the Schiff moment compared to the Ne-TPT interaction is more dominant 
in the CsAg molecule than in FrAg or TlF molecules. Differences in those ratios are crucial in multi-source pictures of beyond SM physics 
since systems with different sensitivity to both interactions lead to stronger constraints on underlying ${\cal{CP}}$-violation
parameters \cite{GaulBerger_JHEP2024}. 
\\
However, the nuclear spin should be included in those ratios. Disregarding details of nuclear structure we can estimate the
nuclear spin-dependent $M$ ratio by a simple approximation, $\left<\Sigma\right>_A \approx I_A$ where $I_A$ is the nuclear angular
momentum quantum number\footnote{A less approximate rendering using angular momentum of an unpaired valence nucleon is presented in 
Ref. \cite{dzuba_flambaum_PRA2009}. The accordingly modified $M$ ratios are not expected to differ qualitatively from the present
values.}. 
The values for $M$ are also displayed in Table \ref{Table:Ratio_SM/T}.
With those three specific isotopes, the ordering according to $M$ is now completely reversed compared to $\tilde{M}$. 
Hence, for these given target isotopes the TlF molecule has a much more sensitive interaction constant for the Schiff moment 
(as compared to the Ne-TPT interaction) than the FrAg molecule.
However, considering the whole picture of both interactions, and not only their interaction constants, changes the perspective and 
interpretation once again.
The Schiff moment itself of $^{223}$Fr is expected to be greater than the Schiff moment of $^{205}$Tl by several orders of 
magnitude\cite{Flambaum_Dzuba_2020,Flambaum_Dzuba_Tan_2020}. 
This difference is mainly explained by the geometry of the nucleus, which the Schiff moment depends on through the octupole and quadrupole deformation parameters\cite{Schiff_scales_as}.
Conversely, even if the Ne-TPT interaction depends on the probed nucleus through spin expectation values of the nucleons 
\cite{Degenkolb_EDMlandscape2024}, these parameters have the same order of magnitude for $^{223}$Fr and $^{205}$Tl \cite{Berengut_Flambaum_Kava_2011,Stadnik_Flambaum_2015}. 
Then we expect this interaction to be roughly the same in both molecules. Finally, considering the entire picture, with interaction constants and interactions themselves, we conclude that experiments using FrAg will give rise to stronger constraints on hadronic $\mathcal{CP}$-violating parameters such as the quantum-chromo-dynamic $\bar{\theta}$ or the $\mathcal{CP}$-violating interaction constants for a coupling between the $\pi$-meson and a nucleon $\bar{g}_n$ ($n=1,2,3$).

All these considerations are established in the framework of a multi-source interpretation of a measured molecular EDM. 
To the contrary, in a single-source scenario assuming the Ne-TPT interaction to make the only contribution to a measured EDM the FrAg molecule is slightly advantaged with regard to other molecules under investigation, like TlF, due to the $^{223}$Fr nuclear spin of 3/2 increasing the $W_T$ interaction constant compared to $^{205}$Tl with a nuclear spin of 1/2. 
Moreover, using laser-cooled atomic gases to assemble a molecule instead of laser-cooling 
molecules as experimental process increases by roughly 3 orders of magnitude the experimental sensitivity (see references in \cite{Fleig_DeMille_2021}). 
All of these considerations predict the FrAg diatomic molecule to be an excellent probe for Ne-TPT interaction and to yield 
constraints on fundamental constants such as the effective coupling constant for semileptonic four-fermion interactions 
$C_{lequ}$ \cite{Dekens_2019}.

\bibliographystyle{unsrt}
\newcommand{\Aa}[0]{Aa}

\end{document}